\documentclass{article}
\usepackage{science}
\usepackage{amsmath,amssymb,amsthm}
\usepackage{epsfig}
\usepackage[mathscr]{euscript} 

\def\urltilde{\kern -.15em\lower .7ex\hbox{\~{}}\kern .04em}
\newtheorem{thm}{Theorem}
\newtheorem{lem}[thm]{Lemma}
\newtheorem{cor}[thm]{Corollary}
\newtheorem{mialc}{Claim}[thm]
\theoremstyle{definition}
\newtheorem{defn}[thm]{Definition}

\DeclareMathOperator{\crank}{r}
\DeclareMathOperator{\tw}{tw}
\DeclareMathOperator{\pw}{pw}
\DeclareMathOperator{\sepn}{s}
\DeclareMathOperator{\ssepn}{\tilde s}
\DeclareMathOperator{\bw}{bw}

\pdfoutput=1

\input supp-pdf.tex
\renewcommand{\epsilon}{\varepsilon}

\begin{document}

\title{On Balanced Separators, Treewidth, and Cycle Rank}
\author
{Hermann Gruber\\
knowledgepark AG, Leonrodstr. 68, D-80636 M\"unchen, Germany}
\date{}
\maketitle

\begin{abstract}
We investigate relations between different width parameters of graphs, 
in particular balanced separator number, treewidth, and cycle rank.
Our main result states that a graph with balanced separator 
number~$k$ has treewidth at least~$k$ but
cycle rank at most~$k \cdot\left(1 + \log \frac{n}{k}\right)$,
thus refining the previously known bounds, 
as stated by Robertson and Seymour~(1986) and by Bodlaender et al.~(1995). 
Furthermore, we show that the improved bounds are best possible.  
\end{abstract}
\section{Preliminaries}
Throughout this paper, $\log x$ denotes the binary logarithm 
of~$x$, and $\ln x$ denotes the natural logarithm of~$x$.

\subsection{Vertex separators}
We assume the reader is familiar with basic notions in graph 
theory, as contained in~\cite{Diestel06}. 
In a connected graph~$G$, a subset $X$ of the 
vertex set~$V(G)$ is a {\em vertex separator} 
for~$G$, if there exists a pair $x,y$ of 
vertices lying in distinct components of~$V(G)-X$, 
or if $V(G)-X$ contains less than two vertices. 
Beside vertex separators, also so-called 
{\em edge separators} are studied in the literature. 
As we shall deal only with the 
former kind of separators, we will mostly 
speak just of {\em separators}  
when referring to vertex separators. 

If~$G$ has several connected components, we 
say~$X$ is a separator for~$G$ if 
it is a  separator for some component 
of~$G$. A  separator~$X$ is {\em (inclusion) minimal} 
if no other separator is properly contained in it.
A separator~$X$ is 
{\em balanced}, if every component of 
$G \setminus X$ contains at most 
$\left\lceil \frac{|V(G)|-|X|}{2} \right\rceil$ 
vertices; and it is called {\em strictly balanced}, if 
every component of $G \setminus X$ contains at most 
$\frac{|V(G)|-|X|}{2}$ 
vertices. 

\begin{lem}\label{lem:exact-separator}
Let $G$ be a graph with a balanced separator of 
size at most~$k$. Then~$G$ has a balanced 
separator of size exactly~$k$.
\end{lem}
\begin{proof}
Assume $X$ is a balanced separator of size~$x<|V(G)|$.
Then we can extend $S$ to a balanced 
separator of size $x+1$ as follows: 
Let $C_1,C_2,\ldots$ be the components of $G-X$, ordered 
by decreasing cardinality.
Let~$v$ be any vertex in the largest component~$C_1$.
We claim that~$X \cup \{v\}$ is again a balanced 
separator for~$G$. Namely, we have 
\[|C_1\setminus \{v\}| \le \left\lceil \frac{|V(G)|-|X|}{2} \right\rceil-1 
\le \left\lceil \frac{|V(G)|-|X\cup \{v\}|}{2} \right\rceil. \]
The components $C_i$ in $G-X$ with $i \ge 2$ are also 
components of $G - (X\cup \{v\})$, so it remains
to bound the cardinality of $C_2$.
The latter is the second largest component of $G-X$, thus  
$|C_2| \le \left\lfloor \frac{|V(G)|-|X|}{2} \right\rfloor$,
and consequently
\[|C_2| \le \left\lfloor \frac{|V(G)|-|X\cup \{v\}| + 1}{2} \right\rfloor 
\le \left\lceil \frac{|V(G)|-|X\cup \{v\}|}{2} \right\rceil. \]
For the components $C_i$ with $i>2$, we have of course
$|C_i|\le |C_2|$, and the proof is completed. 
\end{proof}
 
Arguably, the very first result on graph separators, proved by 
Jordan~\cite{Jordan69} in the $19$th century, 
is most naturally phrased in terms 
of non-strictly balanced separators: 
 every tree admits a balanced
separator consisting of a single vertex. However,
the usage of strictly balanced separators seems 
predominant in the more recent literature 
(see e.g.~\cite{BGHK95,Kloks94,RS86}), 
perhaps for the reason that the use of the ceiling 
operator~$\lceil x \rceil$ seems unappealing. 
But the monotonicity property stated in Lemma~\ref{lem:exact-separator} 
ceases to be true for strictly balanced separators. 
This is already witnessed by very simple examples, 
such as~$P_{2k+1}$, the path graph of odd order~$2k+1$.
In any case, we can always turn a balanced separator 
into a strictly balanced separator by adding at most 
one vertex.

The {\em (strict) balanced separator number} 
of a graph~$G$, denoted by $\sepn(G)$ 
(resp. $\ssepn(G)$ for the strict version), is defined as
the smallest integer~$k$ such that for 
every $Q \subseteq V(G)$, the induced subgraph~$G[Q]$
admits a (strictly) balanced separator of size at most~$k$.
Observe that for any graph~$G$, we have 
$\sepn(G)\le \ssepn(G)\le \sepn(G)+1$.

\subsection{Width parameters}

The {\em cycle rank} of a graph~$G$, denoted by $\crank(G)$, 
is a structural complexity 
measure on graphs and is inductively defined as follows:
If~$G$ has no edges, then $\crank(G) = 1$; if $G$ 
has several components, then $\crank(G)$ is the maximum 
cycle rank among the connected components of~$G$; 
otherwise, $\crank(G) = 1 + \min_{v\in V(G)} \crank(G-v)$.
It easily follows from the definition of cycle rank 
that for every vertex $v\in V(G)$ holds 
$\crank(G - v) \le 1 + \crank(G)$, and similarly 
$\crank(G - X) \le |X|+\crank(G-X)$ for every vertex subset 
$X\subseteq V(G)$.

The notion of cycle rank was originally devised by Eggan 
and B\"uchi~\cite{Eggan63} as a notion on  
digraphs, and appears in the literature 
under many different names, such as 
{\em ordered chromatic number}~\cite{KMS95}, 
{\em vertex ranking number}~\cite{BDJKKMT98}, 
{\em tree-depth}~\cite{NO06}, or
{\em minimum elimination tree height}~\cite{BGHK95,Manne91}. 
Although all these notions ultimately refer to 
the same concept, some sources use a 
different normalization. For instance, the 
minimum elimination tree height 
of a graph is equal to the cycle rank minus~$1$.

Other structural complexity measures studied in this paper 
include the treewidth and pathwidth of graphs; more 
background information on these two measures can be found 
in~\cite{Bodlaender98,Kloks94,Scheffler89}. 
For a graph $G$, let
$\mathscr{V} =\{\,U_1,U_2, \ldots, U_r\,\}$ be a collection
of subsets of~$V(G)$. A tree $\mathscr{T}=(\mathscr{V},\mathscr{E})$ 
with vertex set~$\mathscr{V}$ is called 
a {\em tree decomposition}, if all of the following hold: 
(i) The collection~$\mathscr{V}$ covers the vertex set of the graph~$G$,
in the sense that $V = \bigcup_{U \in \mathscr{V}} U$;
(ii) For every edge $(u,v) \in E$, there is a tree 
node~$U\in\mathscr{V}$ such that both~$u$ and~$v$ are in~$U$; and
(iii) If two tree nodes $U_1$ and $U_2$ are connected in the tree
by a path, then $U_1 \cap U_2$ is a subset of each tree node
visited along this path. 
The {\em width} of a tree 
decomposition $\mathscr{T}=(\mathscr{V},\mathscr{E})$ is defined as 
$\max\{\,|U|-1 \mid U \in \mathscr{V}\,\}$, and the 
{\em treewidth} of~$G$ is defined as the minimum width 
among all tree decompositions for $G$. A {\em path decomposition} is a 
tree decomposition where $\mathscr{T}$ is required to be a path 
graph; and the {\em pathwidth} of~$G$ is the minimum width 
among all path decompositions for $G$.

\section{Main Results}

We recall the following well-known result~\cite[Thm.~11]{BGHK95}, which 
relates the strict balanced separator number~$\ssepn(G)$, 
treewidth~$\tw(G)$, pathwidth~$\pw(G)$, and the cycle rank~$\crank(G)$ of a graph~$G$:

\begin{thm}\label{thm:sep-tw-pw-cr}
Let~$G$ be a graph of order~$n\ge 2$, let~$\ssepn(G)$ denote 
its strict balanced separator number, let $\tw(G)$ and $\pw(G)$ denote 
its treewidth and pathwidth, respectively, 
and let $\crank(G)$ denote its cycle rank. Then
\[\ssepn(G)-1 \le \tw(G) \le \pw(G) \le \crank(G)\le 1 + \ssepn(G) \cdot \log n.\]
\end{thm}

As Bodlaender et al.~\cite{BGHK95} note, the bounds stated in 
this theorem are spread across the literature, and (variations of)
some of these bounds were discovered independently by several groups 
of authors.

Known examples of graphs having a logarithmic gap between 
treewidth and pathwidth are the complete binary trees of 
order~$2^d-1$, which have treewidth~$1$ but pathwidth 
$\lceil d/2 \rceil$, see~\cite{Scheffler89, Scheffler92}. 
For the relation between pathwidth 
and cycle rank, a similar role is played by the path graphs 
of order~$n$, which have pathwidth~$1$ 
but cycle rank~$1 + \lfloor\log n\rfloor$, 
see~\cite{McNaughton69}. Indeed, the cycle rank of 
trees of order~$n$ can be no larger than this~\cite{KMS95}. 
But observe that the strict balanced separator number 
of a path of order~$n$ equals~$2$ for~$n \ge 4$, so 
Theorem~\ref{thm:sep-tw-pw-cr} gives only an upper bound 
of $2\cdot(1+\log n)$
for all trees of diameter at least~$3$. Comparing this with the
upper bound on the cycle rank of trees~\cite{KMS95} noted above, 
we see that the former bound
is {\em a forteriori} unsharp by a factor~$2$. 
The situation is even worse  
for graphs whose balanced separator number is linear in~$n$,
such as complete graphs or expanders: 
their cycle rank can trivially be at most~$n$. 
So the estimate is off by a factor of~$\Omega(\log n)$ in 
this case. Therefore, our first aim will be 
to refine the rightmost inequality of 
Theorem~\ref{thm:sep-tw-pw-cr}. 

The following recurrence will play a crucial role 
in our investigation.
\begin{defn}\label{defn:recurrence}
For integers $k,n\ge 1$, 
let $R_k(n)$ be given by the recurrence
\[R_k(n) = k + R_k\left( \left\lceil \frac{n-k}{2} \right\rceil\right),\]
with $R_k(r_0) = r_0$ for $r_0\le k$.
\end{defn}

As it turns out, the function~$R_k(n)$ can 
serve as upper bound on the cycle rank of
a graph in terms of its order and its balanced separator number:
\begin{lem}\label{lem:sep-cr}
Let $G$ be a graph of order $n$ whose balanced separator number is 
at most~$k$. Then for the cycle rank $\crank(G)$ holds
\[\crank(G) \le R_k(n).\]
\end{lem}
\begin{proof}
The overall structure of the argument is the same 
as e.g. in~\cite{BGHK95,NO06}, but here we 
derive a somewhat stronger statement. 
We prove the statement by induction on the order~$n$ 
of~$G$. The base cases $n \le k$ of the induction 
are easily seen to hold, since the cycle rank of a 
graph is always bounded above by its order.  

For the induction step, assume $n > k$.
Let~$X$ be a balanced 
separator for~$G$ of size exactly~$k$. Using 
Lemma~\ref{lem:exact-separator}, we know that 
such a separator exists. Denote the connected 
components of $G-X$ by $C_1, \ldots, C_p$. 
Then $\crank(G) \le k + \crank(G-X)$, and 
by definition of cycle rank, 
$\crank(G-X) \le \max_{i=1}^p\crank(G[C_i])$. 
As~$X$ is a balanced separator, we have 
$|C_i| \le \lceil\frac{n-k}{2}\rceil$ for 
$1\le i \le p$, so we can apply the induction 
hypothesis to obtain 
$\max_{i=1}^p\crank(G[C_i]) \le R_k(\lceil\frac{n-k}{2}\rceil)$.
Putting these pieces together, we have 
$\crank(G)\le k + R_k(\lceil\frac{n-k}{2}\rceil)$, as desired.
\end{proof}
In fact, we shall see in a moment that this bound is best possible 
for all~$k$ and each~$n \ge k+1$. As in~\cite{Chvatal70},
let $P_n^k$ denote the the $k$th power of 
a path graph of order~$n$, in which two distinct vertices~$u,v$ 
in $V(P_n) = {1,2,\ldots,n}$ are adjacent iff $|v-u| \le k$.
Determining the cycle rank of this graph is
a question recently posed by Novotny et al.~\cite{NON09}
and subsequently answered by Chang et al.~\cite{CKL10}:  

\begin{thm}[Chang et al.]\label{thm:cr-power-path}
Let $k,n\in\mathbb{N}$, and 
let $P_n^k$ denote the $k$th power of a path of order~$n$. 
Then 
\[\crank\!\left(P_n^k\right) = R_k(n).\] 
\end{thm}

Since the function $R_k(n)$ 
from Definition~\ref{defn:recurrence} is arguably very important 
to our context, we will devote more effort to understanding this 
recurrence. Chang et al.~\cite{CKL10} also derive an 
explicit formula, namely
\begin{align}\label{eqn:bulky}
R_k(n) = & \begin{cases}
         n & \mbox{, if } k \ge n-1\\
         k\cdot(\left\lfloor\log\left(1+\frac{n}{k} \right) \right\rfloor-1) +\left\lceil \frac{n+k}{2^{\left\lfloor\log\left(1+\frac{n}{k} \right) \right\rfloor}} \right\rceil & \mbox{, if } k \le n-2 
         \end{cases}
\end{align}
Now we can compute each value of~$R_k(n)$
for given values of $k$ and $n$ effortlessly.
However, the formula is somewhat unwieldy. This calls 
for a more convenient expression for 
reasoning about~$R_k(n)$. Therefore, we shall
derive an easier upper bound in closed form. 
For each fixed~$k$, that upper bound is sharp
infinitely often, so our estimate is essentially 
the best possible. 

\begin{thm}\label{thm:closed-form}
Let $k,n\ge 1$ be integers. Then 
\[R_k(n) \le k \cdot\left(1 + \log \frac{n}{k}\right),\] 
with equality iff $n = k \cdot (2^j-1)$ for some 
$j \in \mathbb{N}\setminus \{0\}$.
\end{thm}
\begin{proof}
Instead of analyzing the recurrence $R_k(\cdot)$ directly, 
we first look at the problem from a different perspective.
To this end, for a given positive integer~$r$, 
let $N_k(r)$ denote the 
smallest integer~$n$ such that $R_k(n) \ge r$.
Then by $R_k(2r+k-1) = R_k(2r+k) = k + R_k(r)$, we have 
$N_k(r+k) = 2 \cdot N_k(r) + k-1$, and the recurrence terminates 
with~$N_k(r_0) = r_0$ for $r_0 \le k$.
 
To get a closed form for $N_k(r)$, we make some use 
of finite calculus (see~\cite{GKP89}): 
The backward difference of $N_k$, denoted by 
$\nabla N_k$, is defined as $\nabla N_k(i) = N_k(i) - N_k(i-1)$. 
For convenience, let us define $N_k(0) = 0$, such 
that $\nabla N_k(1)$ is well-defined.

\medskip

\begin{mialc}\label{claim:delta} For all integers $i,j$ with
$i \ge 1$ and $(i-1)\cdot k< j \le i\cdot k$ holds
\[\nabla N_k(j) = 2^{i-1}.\]
\end{mialc}
\begin{proof}
We prove this by induction on $i$. 
The base cases where $i=1$ and $j\le k$ are easily verified, so it 
remains to perform the induction step. For $(i-1)\cdot k< j \le i\cdot k$, 
we have 
\begin{align}
\nabla N_k(j) & = 2 \cdot N_k(j-k) + k-1-2 \cdot N_k(j-1-k)-k+1 \\
& = 2 \cdot \nabla N_k(j-k) \\
& = 2 \cdot 2^{i-2},
\end{align}
where the last step above holds by 
induction hypothesis, since $(i-2)\cdot k < j-k \le (i-1) \cdot k$. 
This completes the proof of the claim.
\end{proof}

Now that we have a closed form for the backward differences, 
it is not difficult to derive a closed form for $N_k(r)$.

\begin{mialc}\label{claim:closed-form-N}
For all integers $k,r\ge 1$ holds
\[N_k(r) = (k + r \bmod{k}) \cdot 2^{(r - r \bmod{k})/ k} - k.\]
\end{mialc}
\begin{proof}
Telescoping sums yields
$ N_k(r) = N_k(r) - N_k(0) = \sum_{i=1}^r \nabla N_k(i)$. 
Write~$r$ as $r = k \cdot s + t$ with $s = r \operatorname{div} k$ and $t = r \bmod k$. 
We arrive at a simplified form for $N_k(r)$ by grouping 
the summands appropriately, and by rewriting those
using Claim~\ref{claim:delta} afterwards:
\begin{align}
N_k(r) & = 
\sum_{i=1}^{s} \sum_{\begin{subarray}{l}j > (i-1)\cdot k  \\ j\le i \cdot k\end{subarray}} \nabla N_k(j) + \sum_{j= s \cdot k+1}^r \nabla N_k(j)
\\ 
&= \sum_{i=1}^{s}\left(k \cdot2^{i-1}\right) + t \cdot 2^s 
= k \cdot (2^s-1) + t \cdot 2^s.\\
&=  (k + r \bmod{k}) \cdot 2^{(r - r \bmod{k})/ k} - k.
\end{align}

In the last line, we made use of the facts $t = r \bmod{k}$ 
and $s = \frac{r - r \bmod{k}}{k}$.
\end{proof}
Resolving this formula after~$r$ would result in the 
unhandy formula~\eqref{eqn:bulky} by Chang et al.~\cite{CKL10}. 
We thus resort to giving an upper bound that is tight infinitely often. 
  
\begin{mialc}\label{claim:bound} For all integers $k,r$ with $k\ge 2$ and 
$r \ge 1$ holds
\[N_k(r) \ge k \cdot (2^{r/k}-1),\] 
with equality iff $r \bmod{k} = 0$.
\end{mialc}
\begin{proof}
To prove our claim, we 
consider the univariate real-valued function~$f_{k,r}$ given by
\begin{align}\label{eqn:fkrx}
f_{k,r}(x)= (k + x) \cdot 2^{(r - x)/ k} - k,
\end{align} 
with~$k$ and~$r$ positive real numbers greater than~$1$.
Differentiating after~$x$ gives
\begin{align}
f_{k,r}'(x) 
& = 2^{(r-x)/k}\cdot \left(1 - \left(1+ \frac{x}{k}\right)\cdot\ln 2\right). 
\end{align}

An easy computation yields $f_{k,r}'(x) \ge 0$ if and only 
if $x \le k \cdot\left(1 /\ln 2 -1\right)$, 
so the function $f_{k,r}(x)$ has a unique maximum at 
$x_0 = k \cdot\left(1/\ln 2-1\right)$,
and it is monotonically increasing (resp. decreasing) for 
$x < x_0$ (resp. $x > x_0$). 
Now let us restrict the domain of $f_{k,r}$ to 
the closed interval $I = [0;k-1]$. 
Since $0 < 1/\ln 2-1 < 1/2$, the  
number $x_0$ is contained within~$I$ for $k\ge 2$.  

Elementary calculus shows that 
this restricted function will attain its absolute 
minimum at the left or right boundary of~$I$.
Let us calculate which one is the case here:
\begin{align} 
f_{k,r}(k-1) - f_{k,r}(0) &= k \cdot 2^{r/k} \cdot \left(\left(2 - \frac{1}{k}\right)\cdot 2^{(-k+1)/ k} - 1\right).
\end{align} 

By evaluating the right-hand side at $k=1$ and by
differentiating after~$k$, we can deduce that this 
expression is strictly greater than~$0$ for all real-valued $k>1$. Thus, 
$f_{k,r}(k-1) > f_{k,r}(0)$, and we conclude that 
the function $f_{k,r}:I\to \mathbb{R}$ 
attains its absolute minimum at $x_1=0$. 

We return to the function $N_k(r)$, whose domain are the positive 
integers. Recall that Claim~\eqref{claim:closed-form-N} states that
\[N_k(r) = (k + r \bmod{k}) \cdot 2^{(r - r \bmod{k})/ k} - k.\]
Combining this with~\eqref{eqn:fkrx}, we get 
\begin{align}
N_k(r) &= (k + r \bmod{k}) \cdot 2^{(r - r \bmod{k})/ k} - k = f_{k,r}(r \bmod{k}).\end{align} 
We saw above that 
the function $f_{k,r}(\cdot)$, when restricted to the interval $[0;k-1]$, 
attains its 
absolute minimum for $x_1=0$. So we can deduce that 
\[N_k(r) \ge f_{k,r}(0) = k \cdot (2^{r/k}-1)\] 
holds for all~$r\ge 1$ and $k\ge 2$, 
and that equality holds if and only if $r \bmod{k} = 0$.
This completes the proof of the claim.
\end{proof}

We are finally in position to analyze the recurrence $R_k(n)$. 
The special case~$k=1$ is well known and not difficult
to prove by induction (see e.g.~\cite{McNaughton69}), 
so in the following 
we assume~$k\ge 2$. For a given integer~$n$, let $r = R_k(n)$. 
Using the definition of the function $N_k(r)$ and 
Claim~\ref{claim:bound}, we have 
\begin{align}\label{eqn:final-punch}
n \ge N_k(r) \ge k \cdot (2^{r/k}-1), 
\end{align}
with both inequalities being sharp 
iff $n = N_k(r)$ and $r \bmod k = 0$. 
From Claim~\ref{claim:closed-form-N}, we can derive that the latter two conditions are 
in turn equivalent to the requirement that
$n = k \cdot(2^j - 1)$ for some~$j \in \mathbb{N}$.   

With $r = R_k(n)$, we can solve Ineq.~\eqref{eqn:final-punch}
after~$r$ to get $R_k(n) \le k \cdot \left(1 + \log \frac{n}{k} \right)$, with equality 
iff $n = k \cdot(2^j - 1)$ for some~$j \in \mathbb{N}\setminus \{0\}$.
This completes the proof of Theorem~\ref{thm:closed-form}.
\end{proof}

The above result settles the relation between balanced separator number and 
cycle rank. We turn our attention to the other end of the chain 
of inequalities in Theorem~\ref{thm:sep-tw-pw-cr}, namely to
the relation between balanced separator number and treewidth.
Here the optimal inequality will follow by refining a known 
result. 

Recall that a graph is called {\em chordal} iff every 
cycle~$C$ of order greater than~$3$ in $G$ has a {\em chord}, i.e., 
an edge connecting two vertices not adjacent in~$C$. In 
other words, a chordal graph has no induced cycle of 
order greater than~$3$. 

In the following, we derive a 
strengthened version of a theorem originally due 
to Gilbert et al.~\cite{GRE84}---these authors employed a  
looser notion of balanced separator. 

\begin{thm}
Let~$G$ be a chordal graph whose largest clique has order~$p$.
Then~$G$ contains a clique of order at most~$p-1$ that 
is a balanced separator for~$G$. 
\end{thm}
\begin{proof}
The proof follows the one given by 
Gilbert et al.~\cite{GRE84} for a similar statement.
If~$G$ itself is a clique, then removing a single vertex 
yields a balanced separator of order $|V(G)|-1=p-1$. 
So we assume in the following that~$G$ is not a clique.

Choose~$C$ to be the clique in~$G$ which minimizes the 
maximum order among the 
connected components of $G-C$; in case there are several 
such cliques, take~$C$ itself to be of minimum order 
among these cliques. Our aim is to show that then~$C$
is a balanced separator for~$G$. Let~$A$ be a component 
of~$G \setminus C$ that is of maximum order. 
The capstone of the original proof consists of 
deriving the following:

\medskip

{\bf Fact~\cite{GRE84}.} Component $A$ contains 
a vertex~$v$ adjacent to all vertices in~$C$.

\medskip
  
For a proof of this, the reader is referred 
to~\cite[Fact~3]{GRE84}. Notice 
that this fact implies that~$|C| \le p-1$, and thus
it only remains to show that 
$|A| \le \left\lceil \frac{|V(G)|-|C|}{2} \right\rceil$. 
For the sake of contradiction, assume this is not the case.
Then, letting~$B = V(G)\setminus (C \cup A)$, we must 
have $|A|>|B|$, and thus $|B| \le |A|-1$. 
Take a vertex~$v \in A$ that is 
adjacent to all vertices in~$C$. Then $G - (C \cup \{v\})$
falls apart into the disconnected subgraphs~$G[A]-v$
and~$G[B]$, and the maximum order among the connected 
components of $G - (C \cup \{v\})$ is at most~$|A|-1$. 
Thus, the clique $C \cup \{v\}$ would have been preferable 
to~$C$, a contradiction.
\end{proof}

It is well known that if a graph has treewidth at most~$k$, 
then~$G$ is subgraph of some chordal graph whose largest 
clique is of order at most~$k+1$, 
compare~\cite{Bodlaender98,Scheffler89}. The  
following corollary is now immediate:
\begin{cor}\label{cor:sep-tw}
For any graph~$G$, we have~$\sepn(G) \le \tw(G)$. 
\end{cor}

Again, for every possible value of~$\sepn(G)$, there are infinitely 
many graphs for which this bound is tight, as witnessed 
by the $k$th power of a path $P_n^k$. Also, the inequality is 
tight in the case of trees. In this way,
Corollary~\ref{cor:sep-tw} gives a proper generalization of Jordan's
classical result on balanced separators in trees~\cite{Jordan69}.
Since $\ssepn(G)\le \sepn(G)+1$, Corollary~\ref{cor:sep-tw}
also implies the inequality $\ssepn(G)\le \tw(G)+1$ given by 
Robertson and Seymour~\cite[Proposition~2.5]{RS86}. Observe that the latter inequality
is off by~$1$ for all trees containing a path of order~$4$. 

Taking the statements of Lemma~\ref{lem:sep-cr}, 
of Theorem~\ref{thm:closed-form}, and of Corollary~\ref{cor:sep-tw}
together, we have the following improvement over Theorem~\ref{thm:sep-tw-pw-cr}:
\begin{thm}\label{thm:sep-tw-pw-cr-2}
Let~$G$ be a graph of order~$n\ge 2$, and let~$\sepn(G)$ denote 
its balanced separator number, let $\tw(G)$ and $\pw(G)$ denote 
its treewidth and pathwidth, respectively, and let $\crank(G)$ denote its cycle rank. Then
\[\sepn(G) \le \tw(G) \le \pw(G) \le \crank(G)\le \sepn(G) \cdot\left(1 + \log \frac{n}{\sepn(G)}\right).\]
\end{thm}

\section{Applications}

An immediate consequence of Theorem~\ref{thm:sep-tw-pw-cr-2} is a tight upper 
bound for the cycle rank of a graph in terms of its treewidth:
\begin{cor}
Let $G$ be a graph of order $n\ge 2$, and let $\tw(G)$ and $\crank(G)$ denote 
its treewidth and its cycle rank, respectively. Then 
\[\crank(G) \le \tw(G) \cdot\left(1 + \log \frac{n}{\tw(G)}\right).\]  
\end{cor}

Quite obviously, an analogous statement holds for cycle rank versus pathwidth.
Our next application concerns the relation between the cycle rank and 
another width parameter for graphs, namely the bandwidth. The latter is 
defined in the following. 

A {\em linear layout} of a graph~$G$ of order~$n$ is a 
bijection $\ell$ of $V(G)$ into the integer 
interval $[1;n]$. The {\em bandwidth} of a 
linear layout~$\ell$ is defined as $\max_{\{u,v\} \in E(G)} |\ell(v)-\ell(u)|$;
and the bandwidth of $G$, denoted by $\bw(G)$, is defined
as the minimum bandwidth among all linear layouts for~$G$.

The bandwidth of a graph is known to be bounded below by its 
pathwidth, see e.g.~\cite[Thm.~44]{Bodlaender98}, so the 
latter is a 
common lower bound for both cycle rank and bandwidth. 
But how are these two related? It is not difficult to prove
that the bandwidth of the star graph $K_{1,n}$ 
on $n+1$ vertices equals $\lfloor\frac{n+1}{2}\rfloor$, cf.~\cite{Scheffler89}. 
On the other hand, it is clear that $\crank(K_{1,n}) = 2$, so there seems to be 
no interesting upper bound on the bandwidth of a graph in terms of its 
cycle rank. 
For the converse direction, we obtain the following result as a corollary 
to Theorem~\ref{thm:cr-power-path}:
\begin{cor}\label{cor:cr-bw}
Let $G$ be a graph of bandwidth at most~$k$. Then 
\[ \crank(G) \le R_k(n), \]
and this bound is tight if~$G$ is the $k$th power of the 
path graph of order~$n$.
\end{cor}
\begin{proof}
A basic fact about the bandwidth $\bw(G)$ of a graph $G$ of 
order~$n$ is that $\bw(G) \le k$ iff $G$ is isomorphic to a subgraph 
of $P_n^k$~\cite{Chvatal70}. Now we have $\crank(P_n^k) = R_k(n)$, and
the cycle rank can never increase when taking subgraphs.
\end{proof}

As a final application, we determine a sublinear upper bound on 
the cycle rank of the $d$-dimensional hypercube. Recall that the 
vertex set of the $d$-dimensional hypercube is $\{0,1\}^d$, 
and two vertices are adjacent, if their vector representations 
differ in exactly one coordinate. Bounding 
the cycle rank of this graph from below 
has turned out to be useful in formal language 
theory, namely for comparing
the relative succinctness of different variants of the regular 
expression formalism, see~\cite{GH09a}. 
\begin{thm}
Let $H_d$ denote the $d$-dimensional hypercube of order $n = 2^d$. Then 
\[\crank(H_d) = O\left(\frac{n \log \log n}{\sqrt{\log n}}\right).\]
\end{thm} 
\begin{proof}
Harper~\cite{Harper66} proved that for the bandwidth of the $d$-dimensional 
hypercube holds 
\[\bw(H_d) = \sum_{i=0}^d \binom{d}{\lfloor d/2\rfloor}.\]
Using Stirling's approximation, one can show that this sum is 
in~$\Theta(d^{-1/2}2^d)$. To estimate the cycle rank of this graph,
we use Corollary~\ref{cor:cr-bw} and Theorem~\ref{thm:closed-form} 
to obtain:
\begin{align}
\crank(H_d) & \le \bw(H_d) \cdot \log\left(\frac{2^d}{\bw(H_d)}\right)\\
          & = O\left(d^{-1/2}\cdot2^d\cdot\log\frac{2^d}{\Theta\left(d^{-1/2}2^d\right)}\right) \\
          & = O\left(\frac{n \log \log n}{\sqrt{\log n}}\right), 
\end{align} 
as desired.
\end{proof}
As observed in~\cite{GH09a}, the hypercube of order~$n$ 
has cycle rank at least~$\Omega(\frac{n}{\sqrt{\log n}})$, so this upper 
bound is tight up to a factor of~$O(\log \log n)$. In contrast, 
utilizing the known fact that the pathwidth of the hypercube 
is equal to its bandwidth~\cite{CK06} together with the
previously known upper bound from Theorem~\ref{thm:sep-tw-pw-cr} 
would result in an upper bound that even exceeds the trivial upper 
bound of~$n$.

\subsubsection*{Acknowledgement} The author would like to thank Petra Scheffler for sending him a copy of~\cite{Scheffler89} and~\cite{Scheffler92}.

\newcommand\oneletter[1]{#1}

\end{document}